\documentclass{iaus}
\usepackage{graphicx}
\usepackage{natbib}

\title[~~Dust emission in Stephan's Quintet] 
{The dust emission SED of X-ray emitting regions in Stephan's Quintet}

\author[Natale et al.]   
{G. Natale$^1$, R. J. Tuffs$^2$, C. K. Xu$^3$, C. C. Popescu$^1$, J. Fischera$^4$, U. Lisenfeld$^5$, N. Lu$^3$, P. Appleton$^6$, M. Dopita$^7$, P.-A. Duc$^8$, Y. Gao$^9$, W. Reach$^{10}$, J. Sulentic$^{11}$, M. Yun$^{12}$}

\affiliation{$^1$University of Central Lancashire, Preston, PR1 2HE, UK, email: {\tt gnatale@uclan.ac.uk};  
$^2$Max Planck Institute f\"{u}r Kernphysik, Heidelberg, Germany; 
$^3$Infrared Processing and Analysis Center, Caltech, Pasadena, USA; 
$^4$Canadian Institute for Theoretical Astrophysics, University of Toronto, Toronto, Canada; 
$^5$Department de F\'{\i}sica Te\'{o}rica y del Cosmos, Universidad de Granada, Granada, Spain; 
$^6$NASA Herschel Science Center, IPAC, Caltech, Pasadena, USA; 
$^7$Research School of Astronomy \& Astrophysics, The Australian National University, Weston Creek, Australia; 
$^8$Laboratoire AIM, CEA/DSM-CNRS-Universit\'{e} Paris Diderot, Dapnia/Service d'Astrophysique, CEA-Saclay, France; 
$^9$Purple Mountain Observatory, Chinese Academy of Sciences, Nanjing, China; 
$^{10}$Spitzer Science Center, IPAC, Caltech, Pasadena, USA; 
$^{11}$Instituto de Astrof\'{\i}sica de Andaluc\'{\i}a, Granada, Spain; 
$^{12}$Department of Astronomy, University of Massachusetts, Amherst, USA}

\pubyear{2011} 
\volume{284}  
\pagerange{1--12}
\jname{The Spectral Energy Distribution of Galaxies}
\editors{R.J. Tuffs \&  C.C.Popescu, eds.}
\begin{document}

\maketitle

\begin{abstract}
We analysed the Spitzer maps of Stephan's Quintet in order to investigate the nature of the dust emission associated with the X-ray emitting regions of the large 
scale intergalactic shock and of the group halo. This emission can in principle be powered by dust-gas particle collisions, thus providing efficient cooling of the hot gas. 
However the results of our analysis suggest that the dust emission from those regions is mostly powered by photons. Nonetheless dust collisional heating  
could be important in determining the cooling of the IGM gas and the large scale star formation morphology observed in SQ.     

\end{abstract}

\firstsection 
\section{Introduction}

Most of the MIR and FIR dust emission from galaxies is powered by dust absorption of UV/optical photons coming from different 
sources, predominantly young stars in star formation regions, more widely distributed older stellar populations and active galactic nuclei. 
However dust-gas particle collisions can also be important for the dust heating in some cases. 
In particular, for gas densities $\approx 10^{-3}-10^{2}~cm^{-2}$ and gas temperatures $\gtrsim 10^6~K$, a range covering both the shocked interstellar medium (ISM) 
and the group/cluster 
intergalactic medium (IGM), collisionally heated dust temperatures are of order of $\approx10-10^2~K$. 
Therefore, emission from collisionally heated dust is predicted to peak in the MIR/FIR range, as in the case of dust heated by galaxy interstellar radiation fields \citep{P20}.\\
Dust-gas particle collisions are potentially very important for the cooling of hot ISM/IGM gas. 
In fact, for a dust-gas mass ratio higher than $\approx10^{-4}$, the ``dust cooling'' of the hot gas predominates over standard 
bremmstrahlung cooling (see e.g. \citealt{Mo04}). However its efficiency is highly reduced by the rapid removal of dust through 
dust sputtering \citep{Draine79}.\\  
The presence of dust in the hot IGM of groups and clusters has long been searched for, both in direct FIR emission (e.g. \cite{Stickel98}, \cite{Giard2008}, \citealt{Kitayama09})
and through the attenuation of optical light from background galaxies and QSOs (e.g. \citealt{Chelouche07}). 
Although good evidence for grains in the IGM has been found through the extinction technique, it has proven challenging to 
detect IGM dust in the FIR. This is largely due to the need for 
extreme detector stability to measure faint extended structure in well resolved clusters, or, conversely, due to 
confusion with foreground cirrus emission and emission from 
late type member galaxies in more distant clusters with smaller angular extent.  
These problems are alleviated in the case of the two X-ray emitting regions in the Stephan's Quintet compact group of galaxies 
(SQ, distance=$94~Mpc$), the so-called ``shock region'' and the group halo, which are sufficiently resolved to differentiate between 
galaxies and IGM emission while at the same time being not overly extended from the point of view of detector stability. \\
The shock region is an X-ray emitting linear feature of length $\approx 40kpc$ (\citealt{T05}), visible also as $H\alpha$ recombination line (\citealt{Xu99}), radio continuum (\cite{Williams02}), 
warm H2 infrared line emission (\citealt{Cluver10}). Physically, the unusual multi-phase state of the emitting gas has been interpreted as the result 
of a high-speed collision between the ISM of the intruder galaxy NGC7318b and a gaseous tidal tail in SQ IGM (\citealt{Sulentic01}). 
Since these two colliding regions of gas are, in this interpretation, both of galaxian origin, 
they were plausibly dusty before the collision heated them to high temperatures. Collisionally--heated dust is therefore expected in the post-shock gas, if dust sputtering 
has not already completely removed all the dust originally present in the pre-shock gas or if efficient dust injection sources are present. \\ 
As already mentioned, the compactness of SQ also allows an attempt to measure the dust emission associated with the group halo.
Here the gas is expected to be largely pristine in nature, without having been processed in the ISM of a galaxy. However, dust may still
be present if injected into the IGM, for example from halo stars which have been tidally removed from the galaxies, together with 
ISM material. On sufficiently large time scales these stars are expected to become decoupled from the tidally removed ISM through
hydrodynamical interactions between ISM and IGM gas, leaving a population of intergalactic stars in direct contact with the hot IGM. \\
In this proceedings paper we present some of the results of the analysis performed by \cite{Na10} (NA10) on the Spitzer data of SQ. 
A more complete description of the details of this work can be found directly in that paper.   

\begin{figure}[h]
\begin{center}
 \includegraphics[width=5.4in]{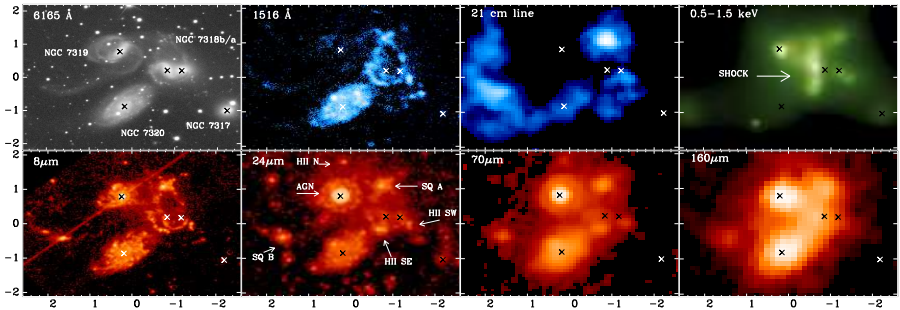} 
 \caption{A multiwavelength view of Stephan's Quintet. Top-panels: SDSS r-band, GALEX FUV, VLA 21 cm line, XMM-NEWTON soft X-ray. 
Bottom-panels: Spitzer IRAC $8~\mu m$ and MIPS $24$, $70$ and $160\mu m$.}
   \label{fig1}
\end{center}
\end{figure}

\section{The dust emission from X-ray emitting regions in Stephan's Quintet}

\begin{figure}[h]
\begin{center}
 \includegraphics[width=3.4in]{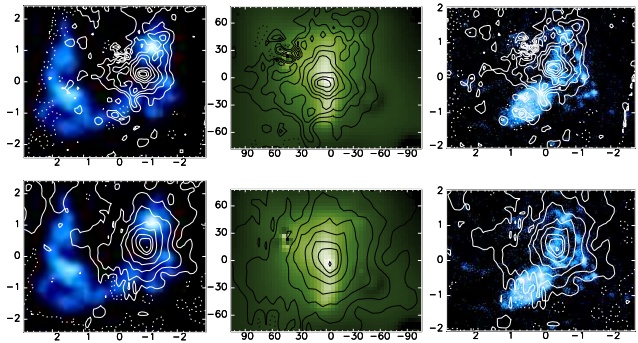} 
 \caption{Contours of the FIR residual maps at $70$ (upper row) and $160~\mu m$ (lower row) overlaid on the HI distribution (VLA, left), the point source subtracted soft X-ray 
map (XMM-Newton, center) and the FUV map (GALEX, right)}
   \label{fig2}
\end{center}
\end{figure}

Fig. \ref{fig1} shows a multiwavelength view of SQ. In the upper row, one can see the stellar distribution, as traced by the SDSS r-band and GALEX FUV maps (note that
NGC7320 is a foreground galaxy), the cold HI distribution (VLA 21cm line) and the soft X-ray emission (XMM-Newton). These maps can be used to compare the stellar and 
cold HI/hot gas distribution with the PAH/dust emission morphology shown in the lower panels, presenting the Spitzer maps at $8\mu m$ (IRAC, stellar light subtracted), 
$24\mu m$, $70\mu m$ and $160\mu m$ (MIPS). An accurate description of the Spitzer maps can be found in NA10. 
Here we simply point out that, by looking at the FIR $70$ and $160\mu m$ maps, the emission from the shock region becomes more predominant and there is a hint for the 
presence of an extended FIR emission spatially coincident with the X-ray halo.\\  
To better understand the origin of the dust emission apparently associated with X-ray emitting regions, we developed a simultaneous multi-source  
fitting technique to remove the sources associated with star formation regions and galaxies in SQ. After the removal of those sources, we obtained FIR residuals maps at 
$70$ and $160\mu m$, which are shown in Fig.\ref{fig2} as contours overlaid on the VLA HI map, the XMM-Newton soft X-ray map (point source subtracted) and the GALEX FUV map. 
The FIR residual emission is dominated by emission from the shock region, but it also presents an extended FIR emission distribution, apparently associated with the halo of 
the group. The multiwavelength comparison in Fig.\ref{fig2} shows that the residual FIR emission is not associated with HI, but is very
 well correlated with X-ray emission instead. This is consistent with the hypothesis of collisionally heated dust emission. However the comparison with the UV map shows 
that there are many UV sources at the same positions, which can potentially heat the dust as well. 
These rather contradictory evidences can be clarified by fitting the dust emission SEDs with dust emission templates, as described in the next section.\\ 

\section{The dust emission SEDs in the shock and halo regions}

\begin{figure}
\begin{minipage}[t]{7cm}
\begin{center}
\includegraphics[width=6cm,clip]{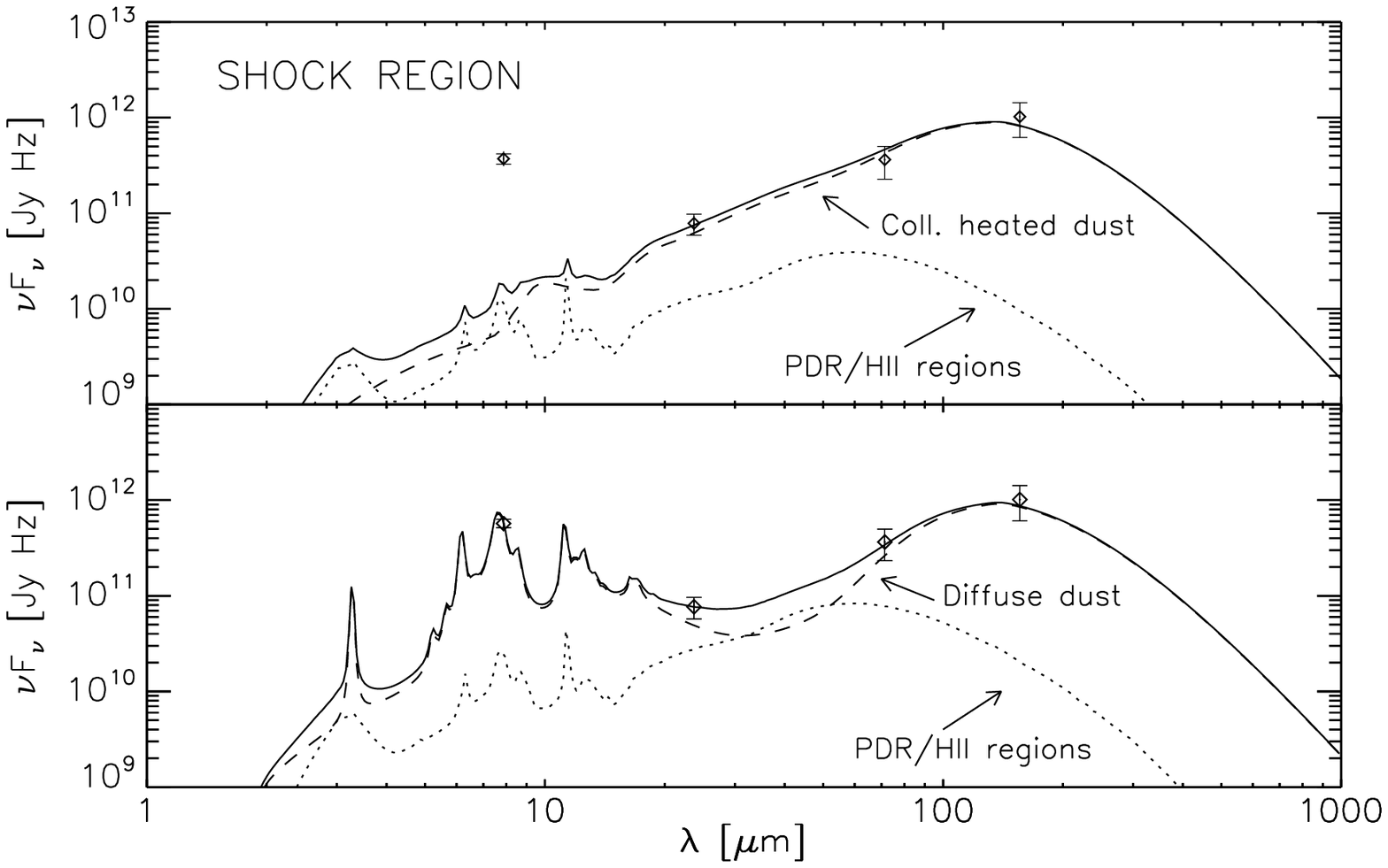}
\caption[Short caption for figure 1]{\label{Fig3} Dust emission SED of the shock region. The upper panel shows the fit using a collisionally heated dust emission template and a star formation region template. The lower panel
shows the fit using a diffuse photon-heated dust emission template and a star formation region template.}
\end{center}
\end{minipage}
\hfill
\begin{minipage}[t]{6cm}
\begin{center}
\includegraphics[width=6cm,clip]{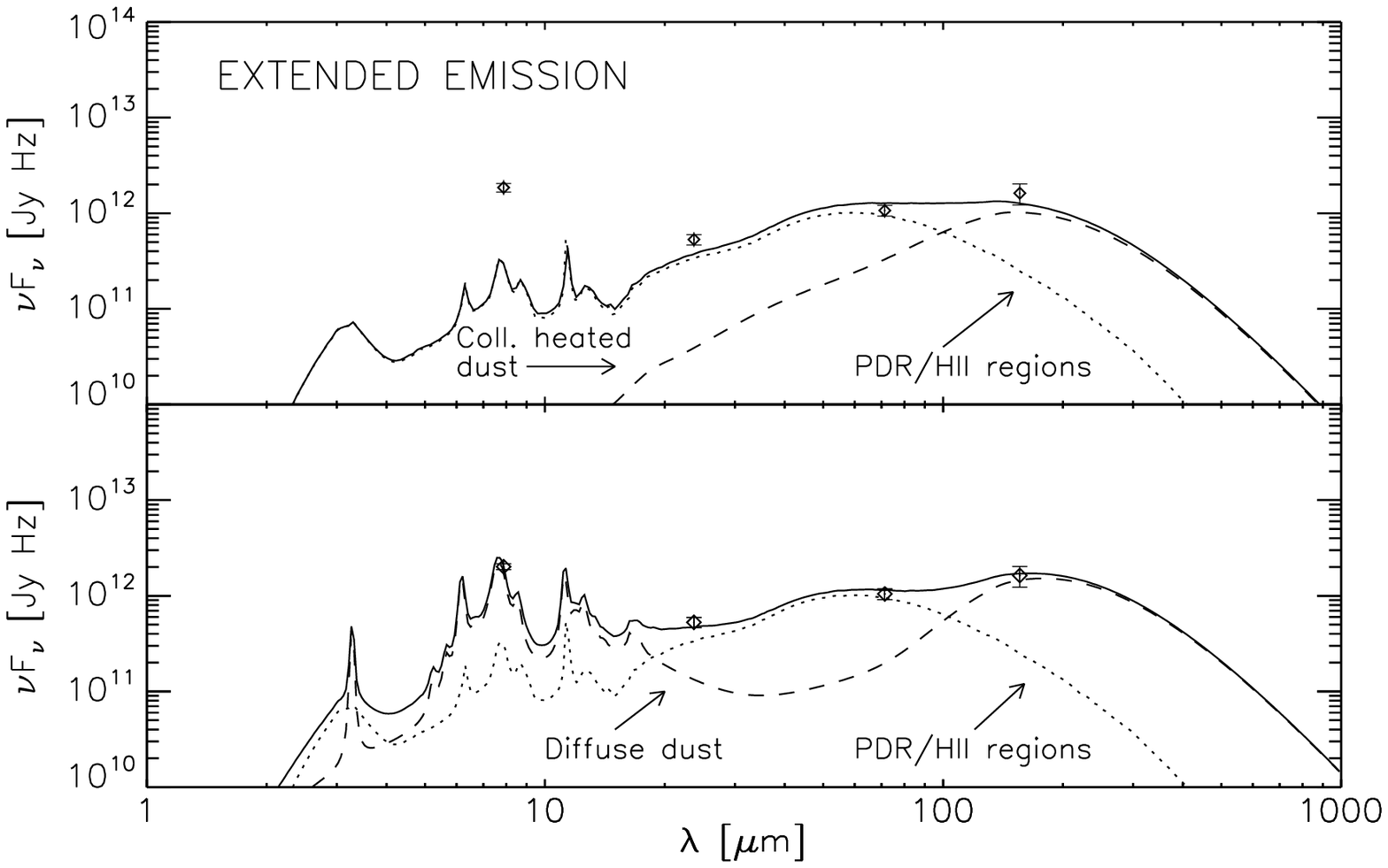}
\caption[Short caption for figure 2]{\label{Fig4} SED of the extended dust emission. The panels show the fits performed with the same kind of templates used 
for the fitting of the shock region dust emission SED (see Fig. 3)}
\end{center}
\end{minipage}
\end{figure}

As shown in NA10, we performed the fit of the dust emission SEDs from the shock and from the halo regions (``the extended FIR emission'') in two ways. 
A first attempt to fit the observed SEDs (see Fig. \ref{Fig3} and \ref{Fig4}, upper panels) has been performed by using a combination of a template for  
collisionally heated dust emission (with plasma parameters fixed to those derived from the X-ray measurements), used to fit mostly the 
FIR, and a template for warm dust emission for star formation regions, used to fit mostly the MIR emission. 
We obtained a good fit for the MIR/FIR continuum ($\lambda \geq 24{\rm \mu m}$), but failed to account for the observed $8{\rm \mu m}$ emission. 
The emission in this band is known to be dominated by 
PAH line emission. These molecules are extremely fragile when embedded in an environment at temperatures of order of $10^6~K$ (\citealt{Micelotta10}). Therefore the relatively 
high level of observed $8{\rm \mu m}$ emission suggests that a large fraction of the emitting dust is actually embedded in a 
cold gas reservoir. In the second attempt, shown in the lower panels of Fig.\ref{Fig3} and \ref{Fig4}, we replaced the template of collisionally 
heated dust emission with a template for a purely photon heated emission component, calculated as a function of combinations of the 
ambient optical and UV radiation fields. We found that, using those templates, one is able to fit very well 
the entire dust emission SEDs. 
These findings, combined with simple estimations of dust injection rates in the hot gas of the shock and of the halo,  
support the interpretation that photon heating of dust has a major role in powering the observed MIR and FIR emission. 
However the few percent of total FIR luminosity 
that can be powered by collisional heating can still be a dominant cooling path for the hot IGM gas (see NA10 for details).  

\section{Conclusions and final remarks}
Our analysis of the Spitzer data of SQ has shown that most of the dust emission associated with the X-ray emitting shock  
and halo regions is heated by photons rather than by gas-dust particle collisions, with a major component of the emitting dust being 
embedded in a cold gas phase (traced by the $8{\rm \mu m}$ PAH dust emission).  
This implies a multi phase medium in which cold infrared emitting clouds are embedded in the hot X-ray emitting plasma. 
In the case of the shock region, one possibility is that the emitting dust (whose mass is ${\rm M_d}\approx 4\times10^7~{\rm M_\odot}$) is mainly in the massive amount of molecular gas recently detected by 
\cite{Guil12}: ${\rm M_{H2}}\approx 4\times10^9~M_\odot$.  
If these cold molecular clouds, presenting low level of star formation, are optically thin then the dust emission can be powered by the external radiation field 
produced by the surrounding stellar populations and the shocked gas. However, 
in an opposite scenario of optically thick clouds, a component of in-situ star formation would be necessary to account for the complete shock 
region SED.\\
In the case of the halo, the presence of widespread
compact MIR sources far away from the galaxy mainbodies suggests that a substantial part of the FIR extended emission we detected is powered by an extended star 
formation pattern in SQ. The large scale correspondence between FIR and X-ray emissions might be explained if there
is distributed star formation fuelled directly by the 
cooling of the X-ray halo. This alternative scenario is potentially very important to understand the fuelling of star 
formation in groups and cluster, and will be further investigated in the context of SQ and other groups.

\end{document}